\documentclass[12pt]{article}
\usepackage{graphics}

\begin{document}
\title{Stable synchronised states of coupled Tchebyscheff maps}
\author{
C. P. Dettmann\thanks{Department of Mathematics, University of Bristol,
University Walk Bristol BS8 1TW, UK}}

\maketitle

\begin{abstract}
Coupled Tchebyscheff maps have recently been introduced to explain
parameters in the standard model of particle physics, using the stochastic
quantisation of Parisi and Wu.  This paper studies dynamical properties of
these maps, finding analytic expressions for a number of periodic states
and determining their linear stability.  Numerical evidence is given for
nonlinear stability of some of these states, and also the presence of
exponentially slow dynamics for some ranges of the parameter.
These results indicate that
a theory of particle physics based on coupled map lattices must specify
strong physical arguments for any choice of initial conditions, and explain
how stochastic quantisation is obtained in the many stable parameter regions.
\end{abstract}

\section{Introduction}
Coupled map lattices~\cite{K93} consist of a continuous variable (here $\Phi$)
defined on a discrete lattice representing space and time.  The dynamics
relates the value of $\Phi$ at each space-time point to its value at
previous lattice points, typically just the most recent time
and the nearest neighbours in space. Coupled map lattices include the
simplest examples of spatiotemporal chaos and are often used to model
less tractable systems such as nonlinear partial differential equations. 

Diffusively coupled Tchebyscheff maps were introduced by Beck in order to
chaotically quantise field theories~\cite{B95}, later applied to spontaneous
symmetry breaking of quantised Higgs fields~\cite{B98}.
The discrete time of the coupled map lattice corresponds to the fictitious
time in the Parisi-Wu stochastic quantisation~\cite{PW}, and is taken
to infinity in order to reproduce quantum mechanics.
The function appearing in the map
is naturally given by the derivative of the quartic double well potential,
which (appropriately scaled) is the third degree Tchebyscheff polynomial
discussed below.  In addition, Tchebyscheff maps are conjugated to
Bernoulli shifts, and so have very chaotic statistical properties, hence
the possibility that the stochastic forces arise intrinsically from the chaotic
dynamics.

In a later work~\cite{B01}, Beck considers coupled map lattices based
on the second degree Tchebyscheff polynomial in addition to the third
degree case considered previously, and restricts attention to one spatial
dimension.  He numerically computes averages of
interaction and self energies, and reproduces to about four digits
a large number of parameters occurring in the standard model of particle
physics, with predictions for some poorly constrained quantities such
as neutrino and Higgs masses.

The opinion of the present author is that although Ref.~\cite{B01} does not
contain a complete theory relating particle physics to coupled map lattices,
the probability of obtaining such a large number of accurate predictions
with so few adjustable parameters is sufficiently small that the models
warrant further attention.  

Quite apart from the context in which these models were introduced, they
have a number of appealing features from a dynamical point of view.  In
particular, there is an adjustable parameter $a$ in which the system takes
two exactly solvable limits, $a=0$ which is decoupled and fully chaotic, and
$a=1$ in which the lattice decouples into two sub-lattices, the latter
of which are fully chaotic for the advanced coupling case and fully stable
for the backward coupling case (see below for definitions).

Motivated by the desire for exact results and the simple form of the equations,
this paper analyses the simplest periodic states, those for which the field
does not depend on the spatial variable, called ``synchronised states''.
Finding these states is equivalent to finding periodic orbits of a one
dimensional map, but the linear stability is a more involved calculation
than one dimensional maps due to spatially dependent unstable modes.
Stability analysis is important because the proliferation of stable states
undermines the argument that stochastic quantisation arises from the
chaotic dynamics.

A general theory of linear stability of periodic solutions of
coupled map lattices is given in Refs.~\cite{AGGN,GA}.  Our analysis
of synchronised states leads, as in these works, to cyclic matrices,
which are always explicitly diagonalisable.  The general theory also permits
spatially periodic solutions of length greater than one (leading to
block cyclic matrices), and lattices of spatial dimension greater than
one (leading to blocks within blocks).   

Finding that an orbit in a spatially extended system is linearly stable
does not imply that a finite measure of initial conditions approaches
the orbit, unlike finite dimensional dynamical systems.  Actually, there
are many possible inequivalent measures of initial conditions: we choose
an initial distribution that is Lebesgue in the field $\Phi$ and
spatially independent, but physical arguments may justify an alternative
choice.  In fact, in view of the results, we will argue that the final
(ie infinite fictitious time) state depends strongly on this choice of
initial conditions, and hence that physical arguments are required to
specify such a choice. 

The linear stability of the synchronised states is given in
Sec.~\ref{s:lin}.  Numerical work in Sec.~\ref{s:non} reveals properties of
the dynamics far from these periodic states. 
The conclusion summarises the results, discussing in more detail the
implications for the Beck theory.

\section{Linear stability}\label{s:lin}
\subsection{Formalism}
As in low dimensional dynamical systems (such as the individual maps),
linear stability analysis in coupled map lattices consists of determining
whether and how fast infinitesimal perturbations of a reference trajectory
(here a periodic orbit) grow with time.  As noted above,
however, linear stability, in that {\em all} sufficiently small
(everywhere) perturbations decay with time, does not imply
that a finite measure of initial conditions is attracted to the reference
trajectory.  Notwithstanding ths caveats of interpretation, the linear
stability analysis is a good starting point for finding stable orbits and their corresponding parameter ranges; nonlinear stability is discussed in
Sec.~\ref{s:non}.

Consider a diffusively coupled map lattice, in which $x$ and $t$ are integer
space and time variables respectively (a more transparent notation than the
conventional $i$ and $n$ respectively):
\begin{equation}\label{e:maps}
\Phi_{x,t+1}=(1-a)f(\Phi_{x,t})
+\frac{a}{2}\left[g(\Phi_{x+1,t})+g(\Phi_{x-1,t})\right]
\end{equation}
Here, $f$ is the map at each site $x$, $g$ couples neighboring maps with a
strength $a$.  Later we will specify $f$ and $g$ to be the possibilities
enumerated in Tab.~\ref{t:maps}.  We demand that $\Phi\in[-1,1]$; this
property is preserved if $f$ and $g$ are closed on this interval, and
$a\in[0,1]$.  Any periodic orbits found by solving the appropriate equations
are only relevant if they lie in this range.

\begin{table}
\begin{tabular}{ccc}\hline
Model&$f$&$g$\\\hline
$2A$&$T_2(\Phi)$&$T_2(\Phi)$\\
$2A^-$&$T_2(\Phi)$&$-T_2(\Phi)$\\
$2B$&$T_2(\Phi)$&$\Phi$\\
$2B^-$&$T_2(\Phi)$&$-\Phi$\\
$3A$&$T_3(\Phi)$&$T_3(\Phi)$\\
$3B$&$T_3(\Phi)$&$\Phi$\\\hline
\end{tabular}
\caption{Definitions of functions $f$ and $g$ appearing in
Eq.~(\protect\ref{e:maps}) for the coupled map lattices considered in this
paper.  The Tchebyscheff polynomials are $T_N(\Phi)=\cos (N\arccos\Phi)$,
specifically $T_2(\Phi)=2\Phi^2-1$ and $T_3(\Phi)=4\Phi^3-3\Phi$.
\label{t:maps}}
\end{table}

Synchronised states are solutions in which $\Phi_{x,t}$ is independent of
$x$, and so are given by the one dimensional map
\begin{equation}
\Phi_{t+1}=(1-a)f(\Phi_t)+ag(\Phi_t)\label{e:sync}
\end{equation}
which is a weighted mean of the $f$ and $g$ maps.  To a first approximation,
stable solutions of the full coupled map lattice which are periodic in time
occur when the synchronised map~(\ref{e:sync}) has a superstable periodic
orbit, that is, when one of its critical points is mapped back to itself
after some time, leading to a Lyapunov exponent of $-\infty$.  We will
observe that the linear stability properties of the full coupled map lattice
are slightly different from that of the synchronised map, so that although
stability occurs in the region of synchronised superstable orbits, the
superstable orbit may not be linearly stable in the extended system.

The difference between stability in the synchronised map and the extended
system is that the full coupled map lattice has an infinite number of
degrees of freedom, leading to a richer spectrum of possible instabilities.
Differentiating Eq.~(\ref{e:maps}) leads to the evolution of the perturbations
\begin{equation}
\delta\Phi_{x,t+1}=(1-a)f'(\Phi_t)\delta\Phi_{x,t}
+\frac{a}{2}\left[g'(\Phi_t)\delta\Phi_{x-1,t}
+g'(\Phi_t)\delta\Phi_{x+1,t}\right]
\end{equation}
or in matrix form
\begin{equation}
\delta\Phi_{t+1}=J_t\delta\Phi_t
\end{equation}
where $J_t$ is a cyclic (assuming periodic boundary conditions)
tridiagonal matrix,
\begin{equation}
J_t=\left(\begin{array}{ccccc}
(1-a)f'(\Phi_t)&\frac{a}{2}g'(\Phi_t)&&&\frac{a}{2}g'(\Phi_t)\\
\frac{a}{2}g'(\Phi_t)&(1-a)f'(\Phi_t)&\frac{a}{2}g'(\Phi_t)&&\\
&\frac{a}{2}g'(\Phi_t)&(1-a)f'(\Phi_t)&\frac{a}{2}g'(\Phi_t)&\\
&&&\ddots&\\
\frac{a}{2}g'(\Phi_t)&&&\frac{a}{2}g'(\Phi_t)&(1-a)f'(\Phi_t)\end{array}
\right)
\end{equation}
The full stability matrix for a periodic orbit of (temporal) length
$\tau$ is a product of the form
\begin{equation}\label{e:jprod}
J=J_{\tau-1} J_{\tau-2}\cdots J_0
\end{equation}

For a cyclic matrix $M_{i,j}=m_{i-j}$ of size $X$ (the full extent of the
system in the spacelike direction), there are $X$ eigenvectors $v_k$, given by
\begin{equation}\label{e:evec}
v_k=(1, e^{{\rm i}k}, e^{{\rm i}2k},\ldots, e^{-{\rm i} k}) 
\end{equation}
where $k=2\pi j/X$, $j=0,1,\ldots,X-1$.  This, together with the fact that
the corresponding eigenvalues $\lambda_k$ are
\begin{equation}
\lambda_k=\sum_{j=0}^{X-1}m_je^{{\rm i}jk}
\end{equation}
follows by substitution into $Mv_k=\lambda_k v_k$.

The product of cyclic matrices $J$ given by~(\ref{e:jprod}) acting on $v_k$
thus gives $\lambda_kv_k$ where the eigenvalue is
\begin{equation}\label{e:stab}
\lambda_k=\prod_{t=0}^{\tau-1}\left[(1-a)f'(\Phi_t)+a g'(\Phi_t)\cos k\right]
\end{equation}
Linear stability follows if $|\lambda_k|<1$ for all values of $k$.
Note that this is a stronger condition than stability of the synchronised
map~(\ref{e:sync}), which is $|\lambda_k|<1$ for $k=0$.

There are two classes of analytic results obtained using Eq.~(\ref{e:stab}),
$\tau\leq2$ in which case the periodic orbits can be given exactly, and
certain families of periodic orbits which are given exactly in the limit
$\tau\rightarrow\infty$.  These are investigated in sections~\ref{s:t=2}
and~\ref{s:long} respectively.

\subsection{Synchronised states of periods one and two}\label{s:t=2} 
The equation for the synchronised dynamics~(\ref{e:sync}) leads directly
to states of period $\tau$ by imposing the condition
\begin{equation}\label{e:per}
\Phi_{t+\tau}=\Phi_t
\end{equation}
For $\tau=2$ it might be supposed that such an equation may
have very complicated, possibly non-closed form solutions, since for the
cubic maps the relevant polynomial is of degree 9 (albeit reducible to degree
6 by noting that the fixed point $\tau=1$ solutions can be removed).  It
turns out that, even in this case,
the solution can be written using arithmetic operations
together with at most two square-root operations.

The periodic orbits are given in Tables~\ref{t:t=1},~\ref{t:t=2}.
We start by noting that maps $2A$ and $3A$, for which the synchronised
maps~(\ref{e:sync}) are just the original Tchebyscheff maps, have no
stable periodic orbits as these are fully expanding when transformed to the
variable $\phi=\arccos\Phi$.  The fixed points ($\tau=1$) of the $2B$ and
$3B$ maps are similar in that they are always unstable, and in that the
solution $\Phi$ does not depend on $a$. The remaining periodic orbits for
$\tau\leq2$ have regions of stability in $a$.

\begin{table}
\begin{tabular}{cccc}\hline
Dynamics&$\Phi$&Domain of validity&Domain of stability\\\hline
$2A$&-1/2&[0,1]&None\\
$2A$&1&[0,1]&None\\
$2A^-$&$\frac{1-\sqrt{9-32a+32a^2}}{4-8a}$&[0,1]&$
(\frac{5}{14}\approx0.3571,\frac{9}{14}\approx0.6429)$\\
$2B$&-1/2&[0,1]&None\\
$2B$&1&[0,1]&None\\
$2B^-$&$\frac{1+a-\sqrt{9-14a+9a^2}}{4-4a}$&[0,1]&
$(\frac{5}{9}\approx0.5556,1)$\\
$3A$&-1&[0,1]&None\\
$3A$&0&[0,1]&None\\
$3A$&1&[0,1]&None\\
$3B$&-1&[0,1]&None\\
$3B$&0&[0,1]&None\\
$3B$&1&[0,1]&None\\\hline
\end{tabular}
\caption{Synchronised states of period $\tau=1$.  The first column
gives the model as defined in Tab.~\protect\ref{t:maps}.  The second column
gives the solution of Eqs.~(\ref{e:sync},\ref{e:per}). The third column
gives values of $a$ for which $\Phi\in[0,1]$.  The fourth column gives
values of $a$ for which $|\lambda_k|<1 \forall k$
in Eq.~(\protect\ref{e:stab}), that is, the orbits are linearly stable.
\label{t:t=1}}
\end{table}

\begin{table}
\begin{tabular}{cccc}\hline
Dynamics&$\Phi$&Domain of validity&Domain of stability\\\hline
$2A$&$\frac{-1\pm\sqrt{5}}{4}$&[0,1]&None\\
$2A^-$&$\frac{-1\pm\sqrt{5-32a+32a^2}}{4-8a}$
&$[0,\frac{4-\sqrt{6}}{8}\approx0.1938]$
&$(\frac{7-2\sqrt{7}}{14}\approx 0.1220,1/6\approx0.1667)$\\
&&$[\frac{4+\sqrt{6}}{8}\approx0.8062,1]$
&$(5/6\approx0.8333,\frac{7+2\sqrt{7}}{14}\approx0.8780)$\\
$2B$&$\frac{-1-a\pm\sqrt{5-18a+9a^2}}{4-4a}$&$[0,1/3\approx0.3333]$
&$(\frac{3-\sqrt{6}}{3}\approx0.1835,\frac{11-\sqrt{96}}{5}\approx0.2404)$\\
$2B^-$&$\frac{-1+a\pm\sqrt{5-14a+9a^2}}{4a-4}$&$[0,5/9\approx0.5556]$
&$(\frac{5-\sqrt{10}}{5}\approx0.3675,5/9\approx0.5556)$\\
$3A$&$\pm\frac{\sqrt{2}}{2}$&[0,1]&None\\
$3A$&$\frac{1\pm\sqrt{5}}{4}$&[0,1]&None\\
$3A$&$-\frac{1\pm\sqrt{5}}{4}$&[0,1]&None\\
$3B$&$\pm\sqrt{\frac{1-2a}{2-2a}}$&$[0,1/2=0.5000]$&$(1/4=0.2500,2/5=0.4000)$\\
$3B$&$\pm\sqrt{\frac{3-4a\pm\sqrt{5-24a+16a^2}}{8-8a}}$
&$([0,1/4=0.2500]$&$[\frac{21-\sqrt{281}}{20}\approx0.2118,1/4=0.2500)$\\
$3B$&$\pm\sqrt{\frac{3-4a\mp\sqrt{5-24a+16a^2}}{8-8a}}$
&$[0,1/4=0.2500]$&$(\frac{21-\sqrt{281}}{20}\approx0.2118,1/4=0.2500)$\\\hline
\end{tabular}
\caption{Synchronised states of period $\tau=2$.  Columns as in
Tab.~\protect\ref{t:t=1} except that the two values of $\Phi$ are represented
by the $\pm$ signs.  The $2A^-$ model has two disjoint domains of validity
of the given orbit.  
\label{t:t=2}}
\end{table}

\begin{figure}
\begin{picture}(290,290)
\put(0,20){\scalebox{0.7}{\includegraphics{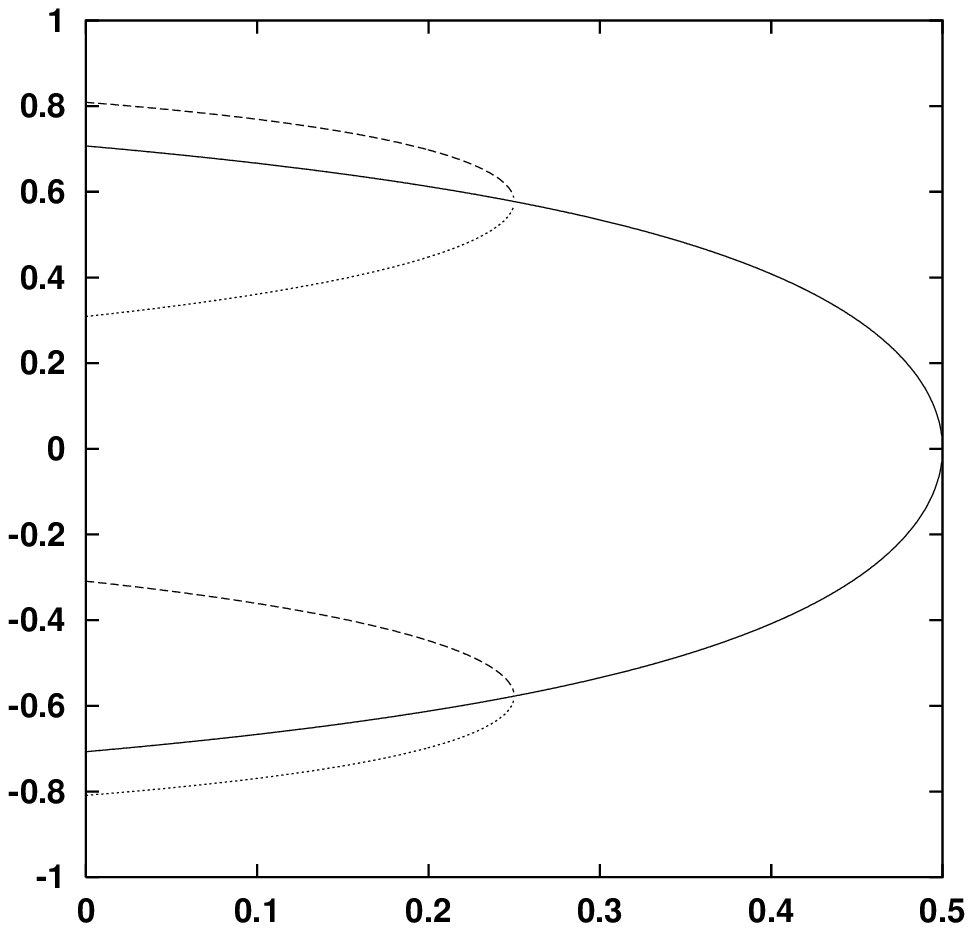}}}
\put(180,0){\Large $a$}
\put(0,180){\Large $\Phi$}
\end{picture}
\caption{Period 2 orbits of the $3B$ model.  The three orbits (see
Tab.~\ref{t:t=2}) are shown using solid, dashed and dotted lines for the
``central'', ``upper'' and ``lower'' orbits respectively; in each
case there are two values of $\Phi$.\label{f:t=2}}
\end{figure}

The calculations here are quite straightforward, so details will be given
for only the most complicated case, the $\tau=2$ orbits of the $3B$ model.
The solutions of the equations~(\ref{e:sync},~\ref{e:per}) are obtained using
a symbolic manipulation package, and are given in Tab.~\ref{t:t=2} and
plotted in Fig.~\ref{f:t=2}, where the three orbits are designated ``central'',
``upper'' and ``lower''.
Equation~(\ref{e:stab}) reads
\begin{equation}\label{e:stab2}
\lambda_k=[3(1-a)(4\Phi_+^2-1)+a\cos k][3(1-a)(4\Phi_-^2-1)+a\cos k]
\end{equation}
where
\begin{equation}
4\Phi_\pm^2-1=\left\{\begin{array}{cc}
\frac{1-2a\pm\sqrt{5-24a+16a^2}}{2-2a}&{\rm upper}\\\\
\frac{1-3a}{1-a}&{\rm central}\\\\
\frac{1-2a\mp\sqrt{5-24a+16a^2}}{2-2a}&{\rm lower}
\end{array}\right.
\end{equation}

The maximum value of $\lambda_k$ is determined from
\begin{equation}
0=\frac{d\lambda_k}{dk}=-a\sin k[3(1-a)(4\Phi_+^2-1+4\Phi_-^2-1)+2a\cos k]
\end{equation}
for which possible solutions are $k=0$, $k=\pi$, and
\begin{equation}\cos k=-\frac{3(1-a)(4\Phi_+^2-1+4\Phi_-^2-1)}{2a}
=\left\{\begin{array}{cc}-3\frac{1-2a}{2a}&{\rm upper/lower}\\\\
-3\frac{1-3a}{a}&{\rm central}\end{array}\right. \end{equation}
These solutions can be substituted back
into Eq.~(\ref{e:stab2}) taking care to exclude ranges of $a$ in which
$\cos k$ is not in $[-1,1]$.

For the central orbit, the $k=0$ solution gives
\begin{equation} \lambda_k=9-48a+64a^2 \label{e:lambeg}\end{equation}
which has magnitude greater than one for $a<1/4$.  The $k=\pi$ solution gives
\begin{equation} \lambda_k=9-60a+100a^2 \end{equation}
which has magnitude greater than one for $a<1/5$ and $a>2/5$.  The solution
$\cos k=-3(1-3a)/a$ is in the range $[-1,1]$ for $a>3/10$ and gives
\begin{equation} \lambda_k=0 \end{equation}
Thus the domain of linear stability for the central orbit is $(1/4,2/5)$ as
given in Tab.~\ref{t:t=2}.

\begin{figure}
\begin{picture}(290,290)
\put(0,20){\scalebox{0.7}{\includegraphics{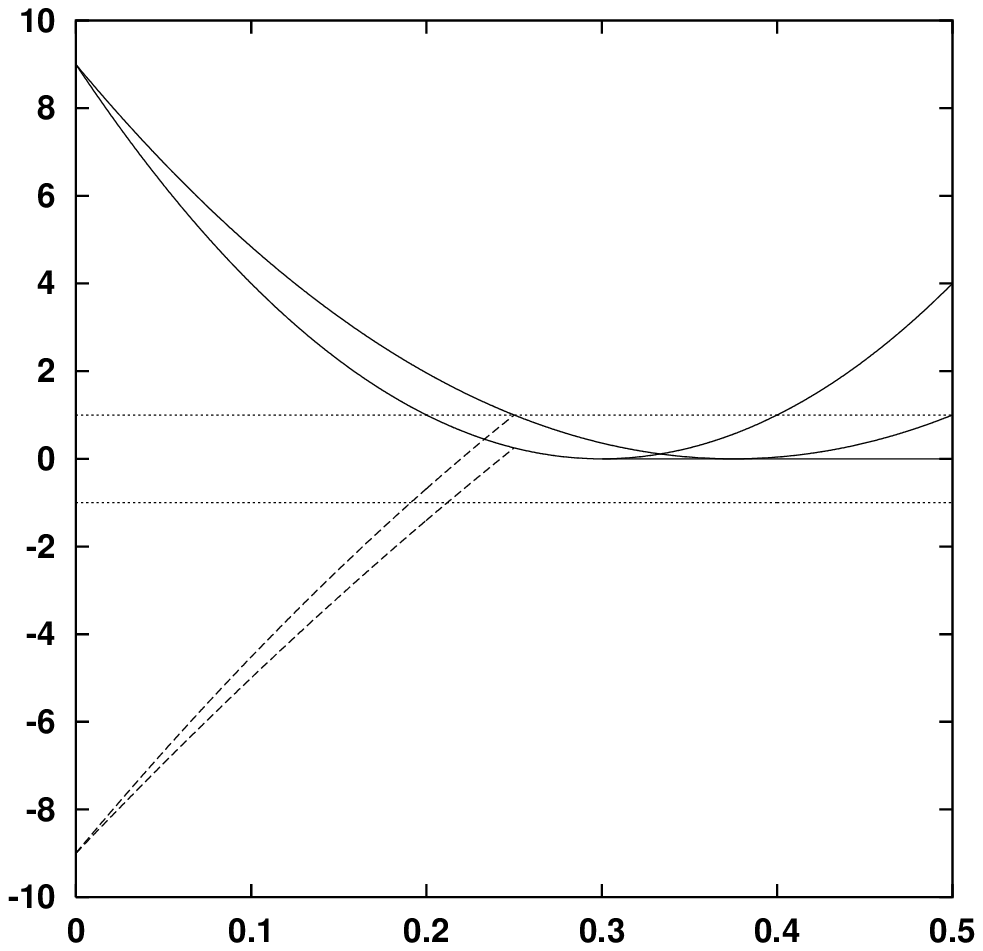}}}
\put(180,0){\Large $a$}
\put(0,180){\Large $\lambda_k$}
\end{picture}
\caption{Linear stability of the orbits shown in Fig.~\protect\ref{f:t=2}.
Depending on the value of $k$, $\lambda_k$ takes values between the solid
or dashed lines for the central or other orbits respectively.  The orbits
are linearly stable if all $\lambda_k$ for a given orbit lie between the
dotted lines.  Formulas for these lines are in
Eqs.~(\ref{e:lambeg}-\ref{e:lamend}).
\label{f:lam2}}
\end{figure}

For the upper and lower orbits, the $k=0$ solution gives
\begin{equation} \lambda_k=-9+48a-32a^2 \end{equation}
which has magnitude greater than one for $a<(3-\sqrt{5})/4\approx0.1910$.
It is also equal to one at the maximum value of $a$ for the orbits, $a=1/4$.
The $k=\pi$ solution gives
\begin{equation} \lambda_k=-9+42a-20a^2 \label{e:lamend}\end{equation}
which has magnitude greater than one for $a<(21-\sqrt{281})/20\approx0.2118$.
The other solution $\cos k=-3(1-2a)/(2a)$ is never in the range $[-1,1]$
for $a\leq 1/4$, and so is irrelevant.  Thus the domain of linear
stability for the upper and lower orbits is $((21-\sqrt{281})/20,1/4)$ as
given in Tab.~\ref{t:t=2}.  The linear stability for all orbits is
presented in Fig.~\ref{f:lam2}.

In this manner all the other domains of linear stability in Tables~\ref{t:t=1}
and~\ref{t:t=2} were generated.  Apart from the $2A$ and $3A$ models for
which the synchronised dynamics is unaffected by the coupling, there are
analytically tractable linearly stable domains of period 2 for all models.
Among the interesting features are the bifurcation observed in the $3B$ model
(see Fig.~\ref{f:t=2}), and two regions of stability in the $2A_-$ model
(see Tab.~\ref{t:t=2}).  We now move to the other analytically tractable case,
the limit $\tau\rightarrow\infty$.

\subsection{Synchronised states of long period}\label{s:long}
This section gives an analytic treatment of families of synchronised stable
orbits with long period.  The idea is similar to that of the superstable
orbits of (for example) the logistic map, in which the critical point (at
which the derivative is zero) maps back to itself after a number of iterations,
leading to a periodic orbit with a Lyapunov exponent of $-\infty$.  Here
the extended nature of the dynamics complicates and destabilises the
dynamics to a degree, however linear stability is still possible, as we
shall show.

\begin{figure}
\begin{picture}(290,290)
\put(0,20){\scalebox{0.7}{\includegraphics{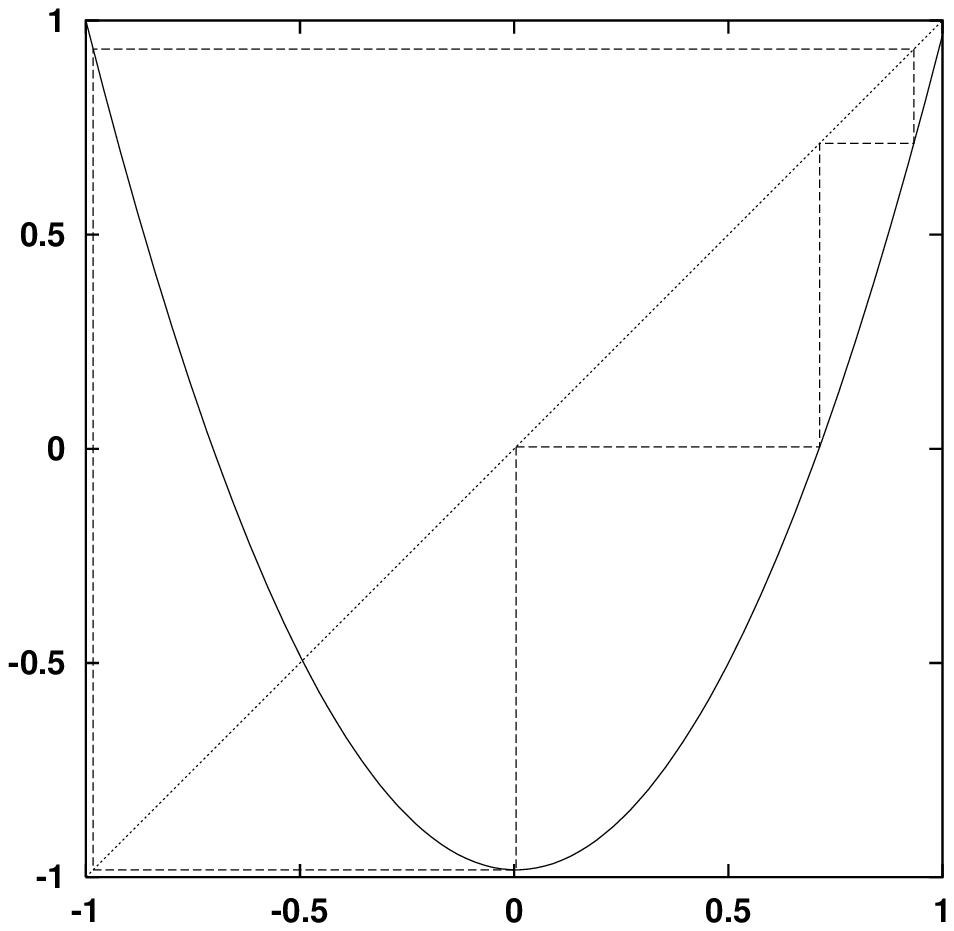}}}
\end{picture}
\caption{Stable $\tau=4$ orbit of the $2B^-$ synchronised map,
Eq.~(\protect\ref{e:sync}).  In this case $a=0.0171$, slightly greater
than the value $a^*/N^\tau=0.0165$ predicted by the large $\tau$ theory
(see Tab.~\protect\ref{t:bn} below).
\label{f:t=4}}
\end{figure}

In the original Tchebyscheff maps the critical points are pre-images of
unstable fixed points at $\Phi=\pm 1$, and so never return to form a
superstable orbit.  However, the synchronised map~(\ref{e:sync}) is, for
small $a$, and excluding the $2A$ and $3A$ cases, a small perturbation on
the Tchebyscheff map, so that after a ``long'' time (of order $-\ln a$) the
trajectory can return to the critical point.  There are many possible
ways in which this can happen; we focus here on the simplest in which
the trajectory, after at most a single point near $\Phi=-1$ always remains
in the rightmost branch of the map.  An example of such a periodic orbit
is given in Fig.~\ref{f:t=4}.

We now locate a periodic orbit of length $\tau$, and in particular calculate
the value of $a$ at which it is close to superstable.  Suppose that for some
value of $a$, the initial point is close to the critical point, that is,
\begin{equation}\Phi_0={\cal O}(a)\end{equation} 
for the second degree maps (that is, $2A^-$, $2B$ and $2B^-$) and
\begin{equation}\Phi_0=-1/2+{\cal O}(a)\end{equation} 
for the third degree map $3B$.  We can use Eq.~(\ref{e:sync}) to find the
successive iterates.  We have
\begin{equation}\Phi_1=-1+b_1a+{\cal O}(a^2)\label{e:b1}\end{equation}
for the second degree maps and
\begin{equation}\Phi_1=1-b_1a+{\cal O}(a^2)\end{equation}
for the third degree map.  Here $b_1$ is a known constant that depends on the
map, and is given in Tab.~\ref{t:bn}.  Iterating again, we have
\begin{equation}\Phi_2=1-b_2a+{\cal O}(a^2)\end{equation}
for all maps.  Again, $b_2$ is given in Tab.~\ref{t:bn} for each model.

\begin{table}
\begin{tabular}{ccccccccc}\hline
Model&$b_1$&$b_2$&$d$&$b^*$&$a^*$&$\pi^*$&$\lambda^*$\\\hline
$2A_-$&2&10&2&2/3&$3\pi^2/16\approx1.851$&$1/(2\pi)$&0\\
$2B$&1&6&0&3/8&$\pi^2/3\approx3.290$&$1/(2\pi)$&$\pi/6\approx0.5236$\\
$2B_-$&1&4&2&7/24&$3\pi^2/7\approx4.230$&$1/(2\pi)$&$3\pi/14\approx0.6732$\\
$3B$&3/2&27/2&0&1/6&$4\pi^2/3\approx13.159$&$1/(4\pi\sqrt{3})$&
$\pi/(3\sqrt{3})\approx0.6046$\\\hline
\end{tabular}
\caption{Coefficients of the long period orbits, defined throughout
Sec.~\protect\ref{s:long}.\label{t:bn}}
\end{table}
 
Further iterates remain near the unstable fixed point of the unperturbed
map at $\Phi=1$, that is, we have
\begin{equation}
\Phi_j=1-b_ja+{\cal O}(a^2)
\end{equation}
where each successive $b_j$ is obtained from the previous one by an equation
of the form
\begin{equation} b_{j+1}=N^2b_j+d\label{e:rec} \end{equation}
where $N$ is the degree of the map (two or three) and $d$ is given in
Tab.~\ref{t:bn} for each map.
The recursion~(\ref{e:rec}) is a linear difference equation,
solved by multiplying through by $N^{-2(j+1)}$ and summing over $j$ from $2$ to
$m-1$.  The result is
\begin{equation}
b_n=N^{2(m-2)}b_2+d\frac{N^{2(m-2)}-1}{N^2-1}\approx N^{2m}b^*
\end{equation} 
where
\begin{equation}
b^*=\frac{b_2}{N^4}+\frac{d}{N^6-N^4}
\end{equation}
is also given in Tab.~\ref{t:bn}.

As the trajectory approaches the end of the periodic orbit, $\Phi$ deviates
substantially from $1$, invalidating the above approximation.  This part
of the trajectory is obtained by iterating $\Phi_0=\Phi_\tau$ backwards.
Recalling that the Tchebyscheff polynomials can be represented as
$T_n(\Phi)=\cos n\arccos\Phi$ we have for the degree 2 models
\begin{eqnarray}
&&\Phi_{\tau-1}=\cos\frac{\pi}{4}+{\cal O}(a)=\sqrt{\frac{1}{2}}+{\cal O}(a)\\
&&\Phi_{\tau-2}=\cos\frac{\pi}{8}+{\cal O}(a)=
\sqrt{\frac{1}{2}\left(1+\sqrt{\frac{1}{2}}\right)}+{\cal O}(a)\\
&&\Phi_{\tau-n}=\cos\frac{\pi}{2^{n+1}}+{\cal O}(4^{-n}a)=
\sqrt{\frac{1}{2}\left(1+\sqrt{\frac{1}{2}\left(1+\ldots\right)}\right)}
+{\cal O}(4^{-n}a)
\end{eqnarray}
and for the degree 3 model
\begin{equation}
\Phi_{\tau-n}=\cos\frac{2\pi}{3^{n+1}}+{\cal O}(9^{-n}a)
\end{equation}
Note that the ${\cal O}(a)$ corrections are suppressed by the contraction
of the inverse map.

The orbit is closed by matching the expressions for $\Phi_m$ and
$\Phi_{\tau-n}$, assuming that both $m$ and $n$ (actually $N^{2m}$ and $N^{2n}$)
are much greater than one.  The result is
\begin{equation}
1-b^*4^{\tau-n}a+{\cal O}(4^{-n}a)
=1-\frac{1}{2}\left(\frac{\pi}{2^{n+1}}\right)^2
+{\cal O}(4^{-2n})
\end{equation}
for the second degree models and
\begin{equation}
1-b^*9^{\tau-n}a+{\cal O}(9^{-n}a)
=1-\frac{1}{2}\left(\frac{2\pi}{3^{n+1}}\right)^2
+{\cal O}(9^{-2n})
\end{equation}
for the third degree model.  The value of $a$ can now be read off as
\begin{equation}
a=a^*N^{-2\tau}+{\cal O}(N^{-4\tau})
\end{equation}
where $a^*$ is given in Tab.~\ref{t:bn}.  The result is consistent with the
original assumption that $a$ is small.  Note that the original ${\cal O}(a)$
freedom in the initial condition $\Phi_0$ contributes to ${\cal O}(a^2)$
when the
trajectory is near $\Phi=1$, and so is expected to shift $a$ by an amount
of order $a^2$, that is, $N^{-4\tau}$.  Obtaining the appropriate
coefficient would require a more involved calculation.

Now we turn to the linear stability of these long orbits using
Eq.~(\ref{e:stab}).  $\lambda_k$ is now a product of a large number
($\tau$) of factors.  The $t=0$ factor is small since both $f'(\Phi_0)$ and
$a$ are small, and will be considered separately.  For the remaining factors
the $f'(\Phi)$ term dominates and the $g'(\Phi)$ term can be ignored, owing
to the presence of the small quantity $a$.  Thus we have
\begin{equation}
\lambda_k=[f'(\Phi_0)+ag'(\Phi_0)\cos k]\prod_{t=1}^{\tau-1}f'(\Phi_t)
[1+{\cal O}(a\tau)]
\end{equation}

The product is
\begin{equation}
\prod_{t=1}^{\tau-1}f'(\Phi_t)=
-\prod_{t=1}^{\tau-1}4\cos\frac{\pi}{2^{\tau-t+1}}
=-\prod_{t=1}^{\tau-1}2\frac{\sin\frac{\pi}{2^{\tau-t}}}
{\sin\frac{\pi}{2^{\tau-t+1}}}
=-2^{\tau-1}\frac{\sin\frac{\pi}{2}}{\sin\frac{\pi}{2^{\tau}}}
\end{equation}
for the second degree models, ignoring corrections of order $a$. An analogous
result holds for the third degree model.  Making
a small angle approximation we arrive at
\begin{equation}
\left|\prod_{t=1}^{\tau-1}f'(\Phi_t)\right|=N^{2\tau}\pi^*
\end{equation}
where $\pi^*$ is given in Tab.~\ref{t:bn}.  Thus the product,
being of order $N^{2\tau}$, balances the first factor
which is of order $a$ and hence $N^{-2\tau}$.  An orbit will be linearly
stable if the remaining coefficient maximised over $k$ (and denoted
$\lambda^*$) is less than one.

We can set $f'(\Phi_0)=0$ by letting $\Phi_0=0$ for the second degree models
and $\Phi_0=-1/2$ for the $3B$ model.  For the $2A^-$ model we
also have $g'(0)=0$ so the orbit remains superstable in the presence of
perturbations at all wavenumbers $k$, and $\lambda^*=0$.  The $B$ models
have $g'(0)=\pm1$, so that we have stability if $\lambda^*=a^*\pi^*<1$.
This quantity is also given in Tab.~\ref{t:bn}, and it shows that all
orbits are linearly stable when $f'(\Phi_0)=0$.

Note that the most stable state is when $f'(\Phi)=0$ belongs to a periodic
orbit of the synchronised map~(\ref{e:sync}); this is not generally
the same as the superstable point of this map, which is the solution of
the equation $(1-a)f'(\Phi)+ag'(\Phi)=0$.

Finally we make a connection to an observation made in Ref.~\cite{B01}.
In section 6.6 of that work a scaling was observed numerically in the limit
$a\rightarrow0$, in particular the dynamical average
\begin{equation}\label{e:V}
V(\Phi)=\langle\int_0^\Phi f(\Phi')d\Phi'\rangle
 \end{equation}
exhibited the following behaviour
\begin{equation}
V(a)-V(0)=h(a)\sqrt{a}
\end{equation}
where the function $h(a)$ satisfies (in the limit)
\begin{equation}
h(N^2a)=h(a)
\end{equation}
that is, $h(a)$ is log-periodic.  Here we have found a family of stable
orbits in which $a$ differs from one to the next by a factor of $N^2$.  The
$N^2$ comes from the derivative of the synchronised map at the points
$\Phi=\pm 1$.  The scaling of $N^2$ would occur in other families of orbits
(stable or unstable) containing $\Phi=\pm 1$, and hence possibly to all
small values of $a$.  

\section{Nonlinear stability}\label{s:non}
Linear stability, considered in the previous section, implies that sufficiently
small perturbations of a periodic state will approach that state in the future.
However, this does not imply that a generic initial state will approach the
stable state.  For a sufficiently large system, some of the initial conditions
will be near points of the periodic state, however there is no guarantee that
neighbouring maps will not strongly perturb these maps away from the periodic
state.

This section contains numerical results that use an initial state in which
the $\Phi_{x,0}$ initial conditions are independently distributed uniformly
from the interval $[-1,1]$.  The spatial size of the system is $10^3$ maps
with periodic boundary conditions.  After a time of $10^4$ units
the values of $\Phi$ for each of the maps are plotted for many values of the
parameter $a$.  The results do not depend noticeably on the spatial extent,
but do depend on the relaxation time (see below).
See Fig.~\ref{f:non}.

\begin{figure}
\begin{picture}(400,560)
\put(10,420){\scalebox{0.45}{\includegraphics{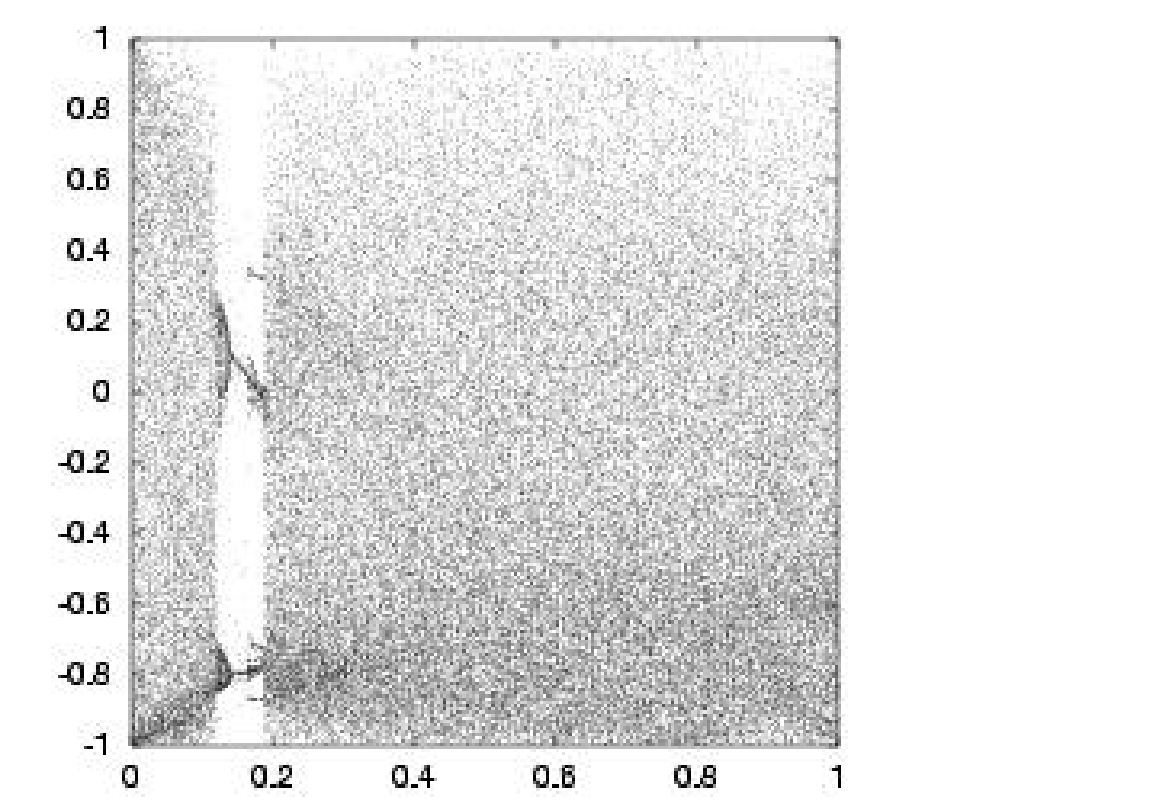}}}
\put(220,420){\scalebox{0.4}{\includegraphics{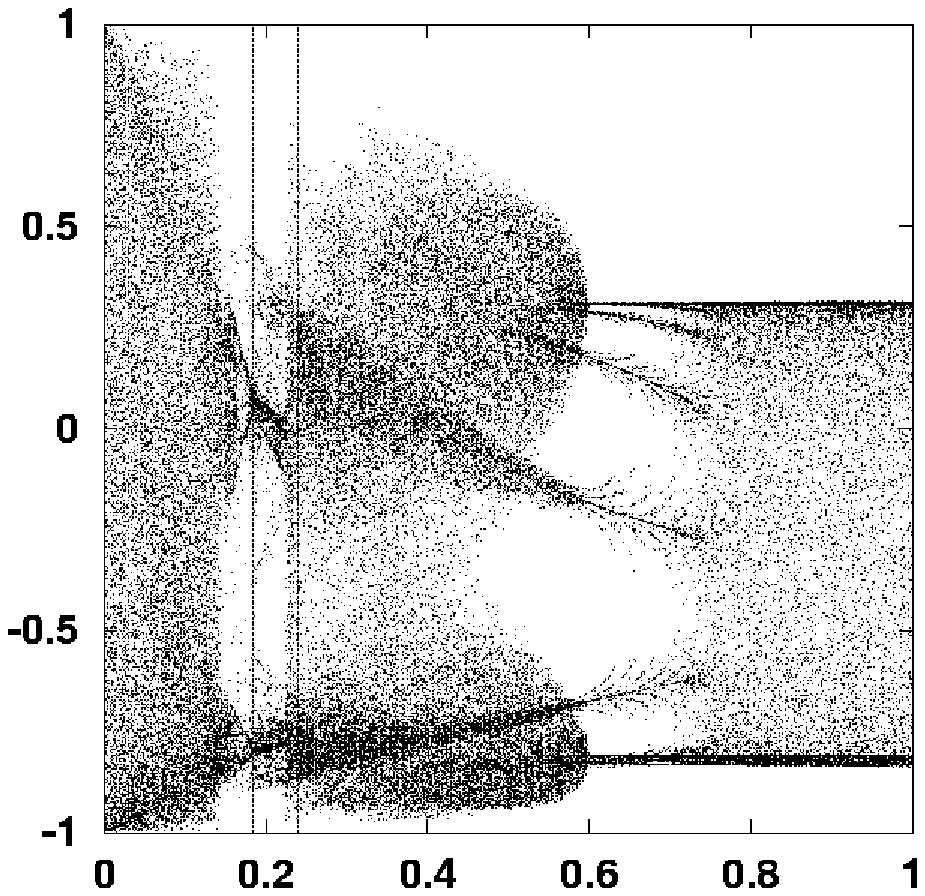}}}
\put(20,220){\scalebox{0.4}{\includegraphics{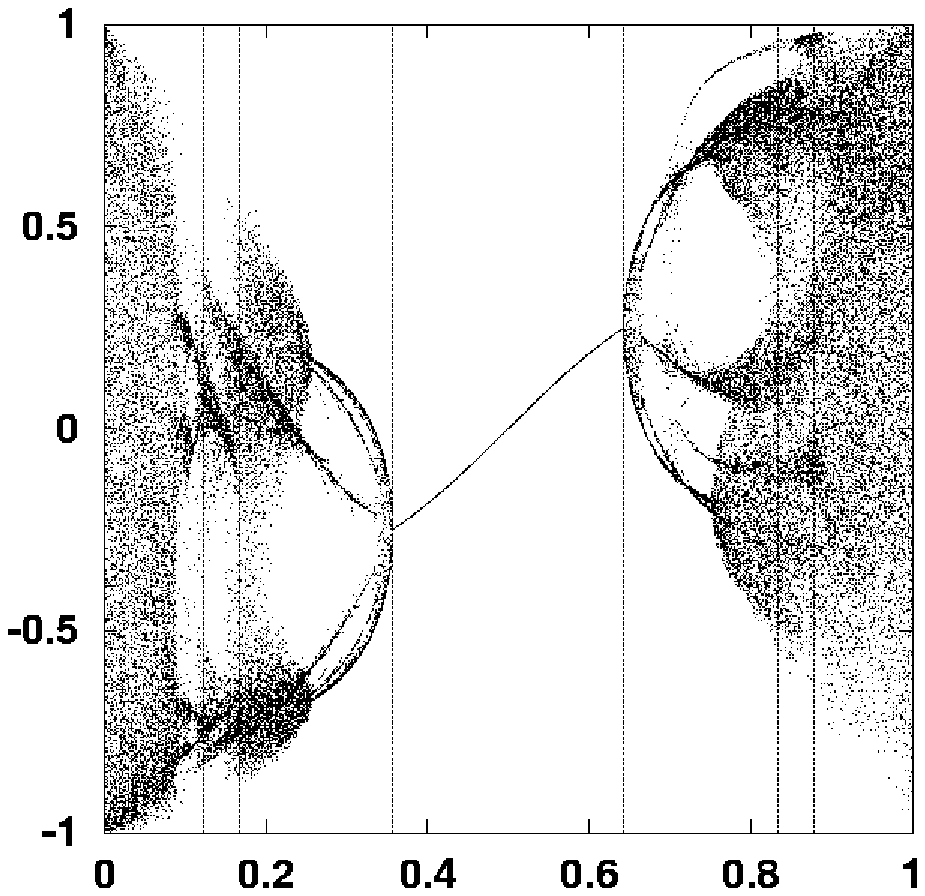}}}
\put(220,220){\scalebox{0.4}{\includegraphics{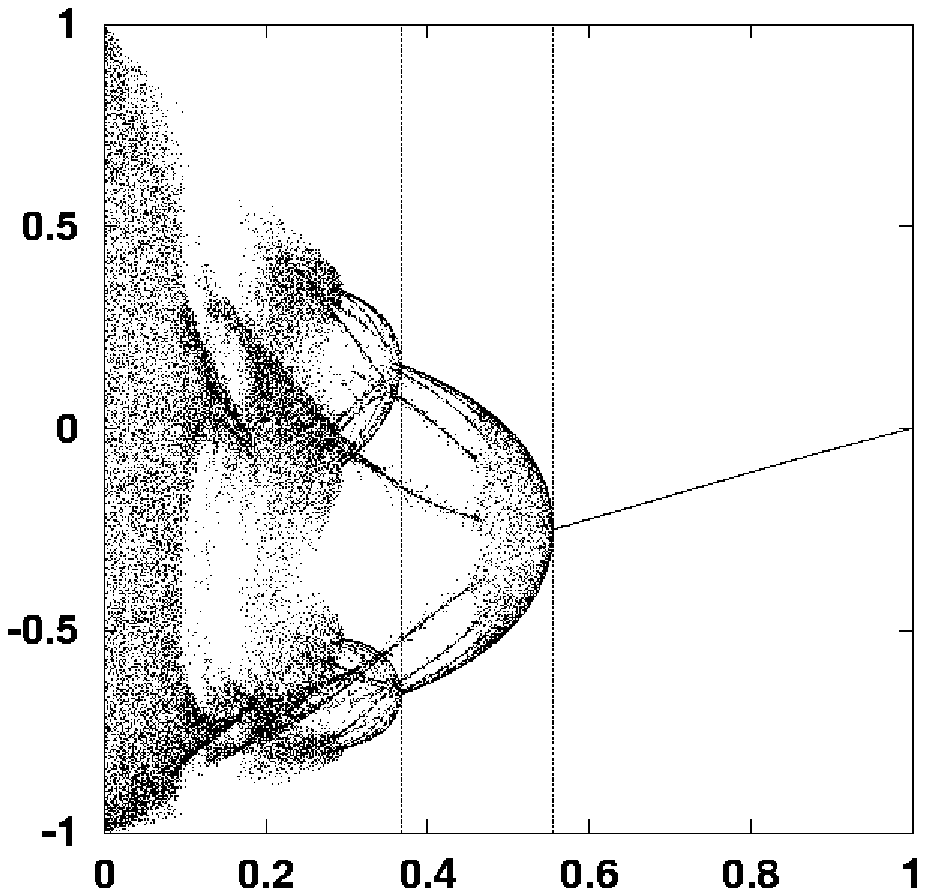}}}
\put(20,20){\scalebox{0.4}{\includegraphics{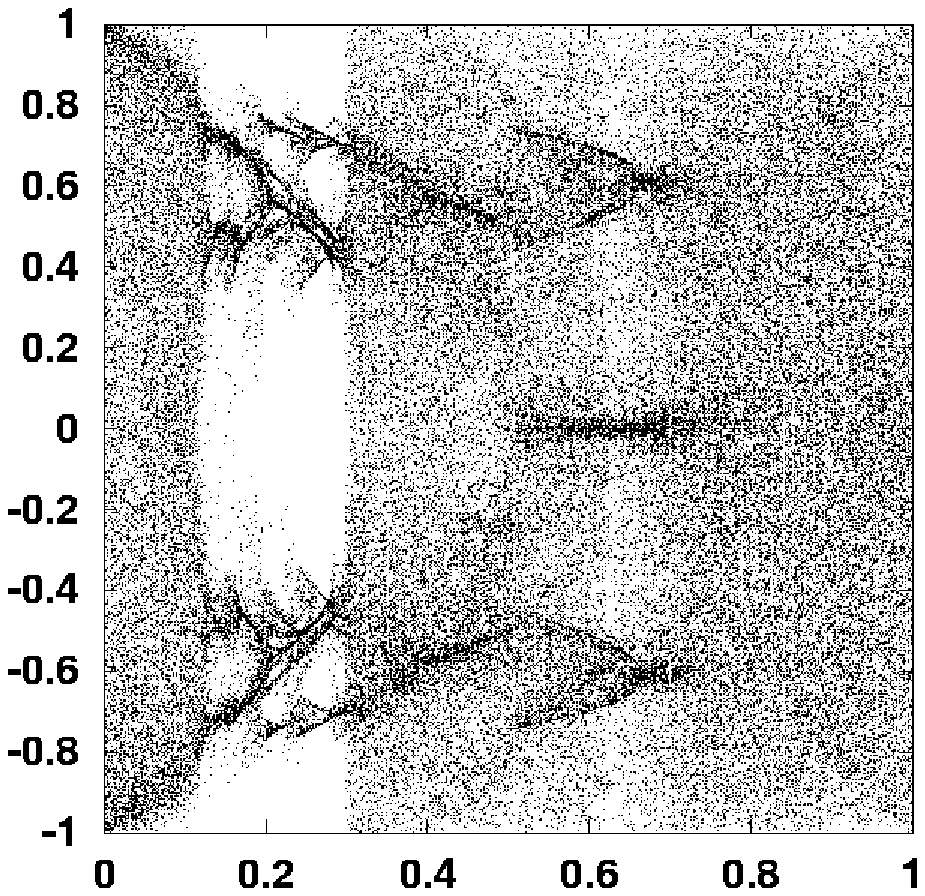}}}
\put(220,20){\scalebox{0.4}{\includegraphics{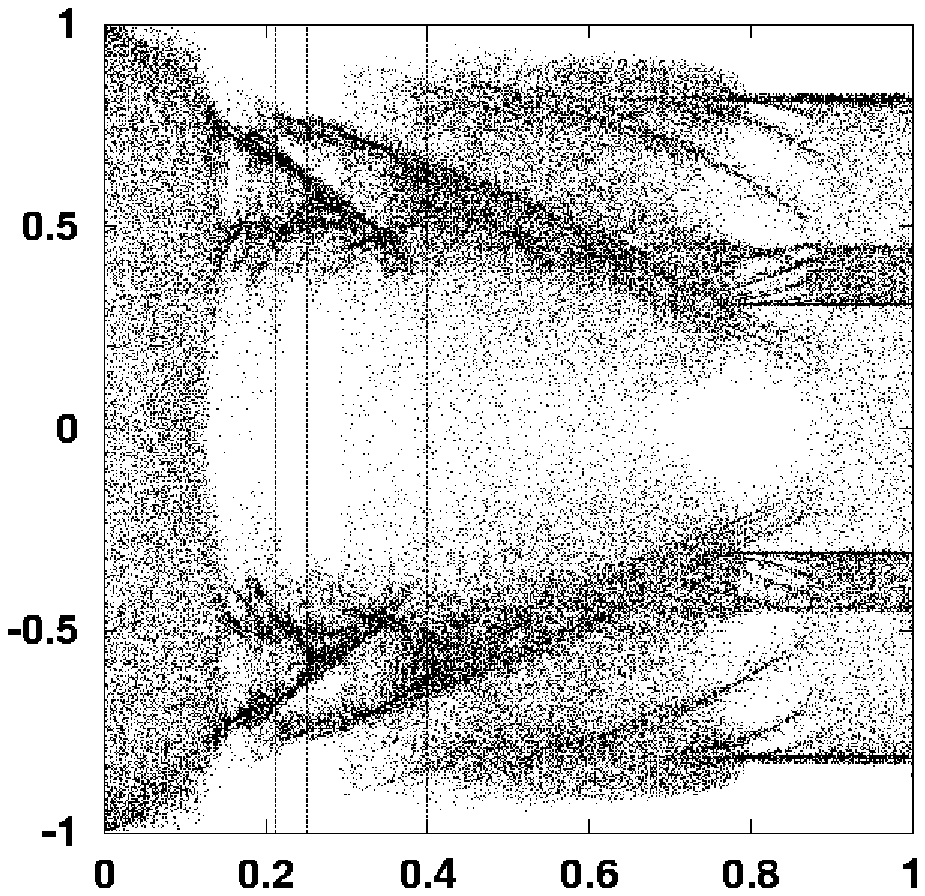}}}
\put(100,0){\Large $a$}
\put(0,100){\Large $\Phi$}
\end{picture}
\caption{Bifurcation diagrams for each of the models. The left column from
top are $2A$, $2A^-$ and $3A$ respectively; the right column is the same
with $A$ replaced by $B$.  The vertical lines indicate boundaries of
linearly stable orbits from Tabs.~\protect\ref{t:t=1},~\protect\ref{t:t=2}.
\label{f:non}}
\end{figure}

The boundaries of the linearly stable orbits of period 1 and 2 given in
Tabs.~\ref{t:t=1} and~\ref{t:t=2} are shown as vertical lines; most of
these boundaries are clearly bifurcations.  The period 1 orbits are both
stable; the period 2 orbits while having a clearly observable effect on
the dynamics, do not appear to be the entire attractor; the high period
orbits discussed in Sec.~\ref{s:long} appear completely absent, although
they are visible if the initial conditions are chosen closer to the
periodic state.

\begin{figure}
\begin{picture}(400,560)
\put(0,420){\scalebox{0.4}{\includegraphics{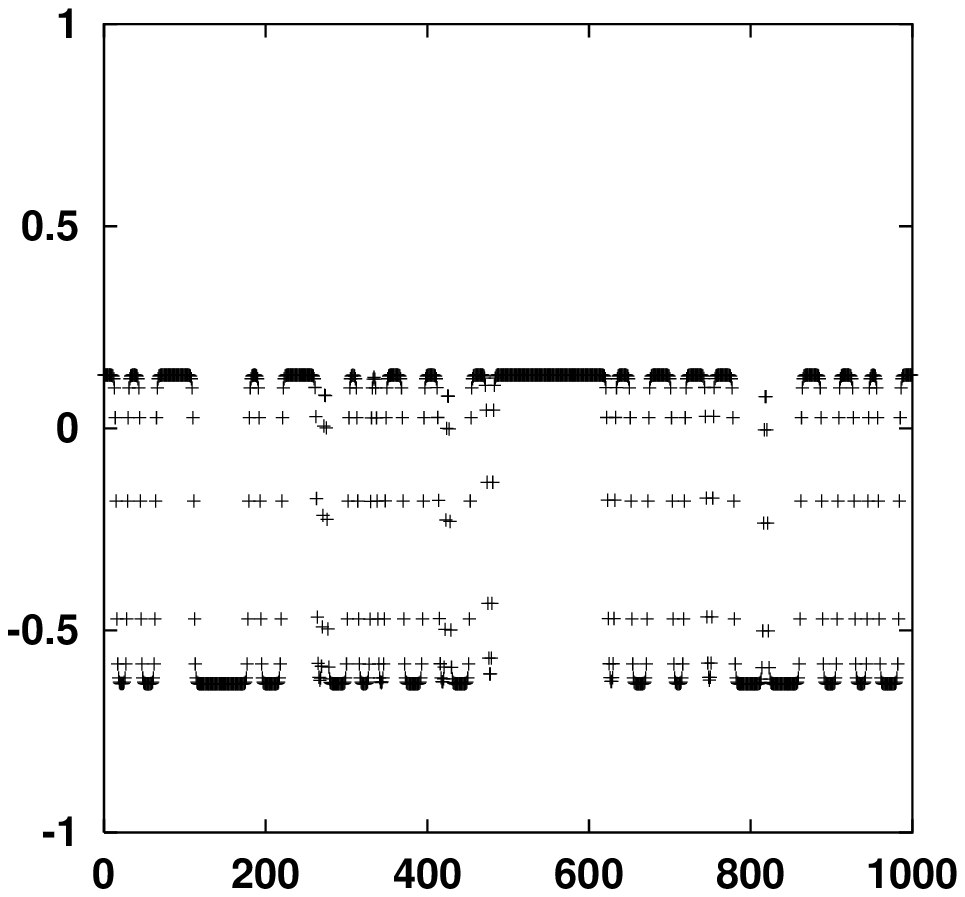}}}
\put(200,420){\scalebox{0.4}{\includegraphics{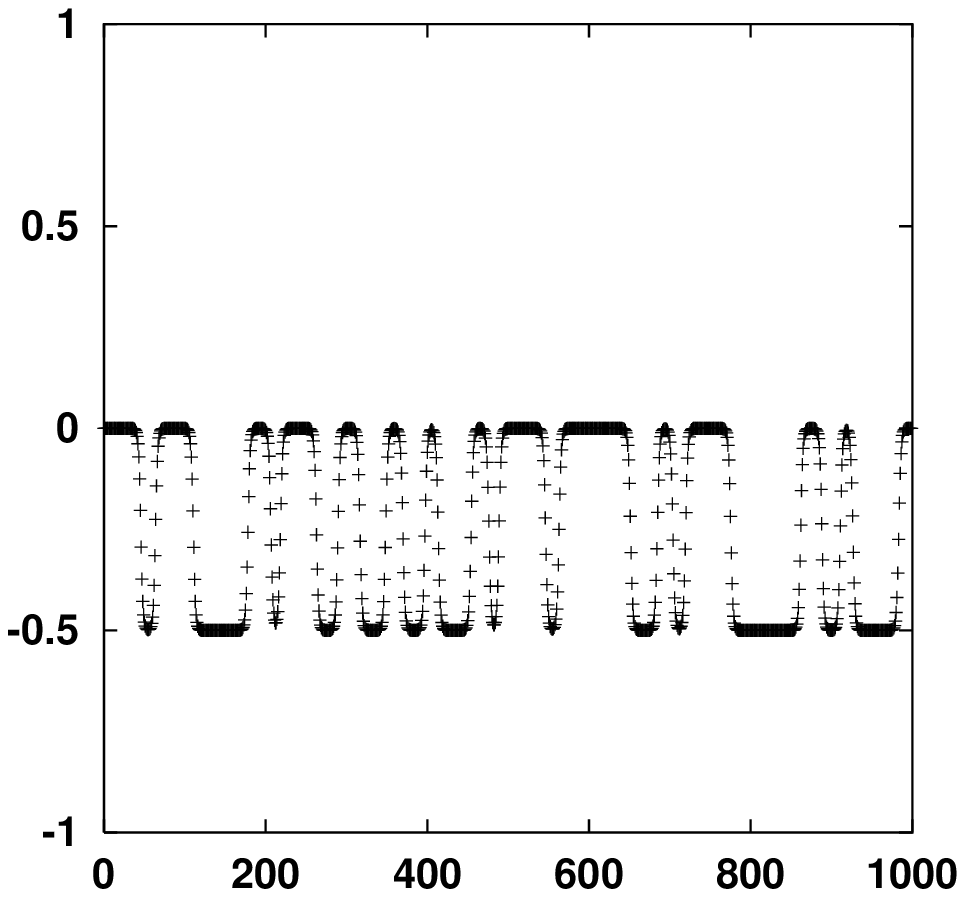}}}
\put(0,220){\scalebox{0.4}{\includegraphics{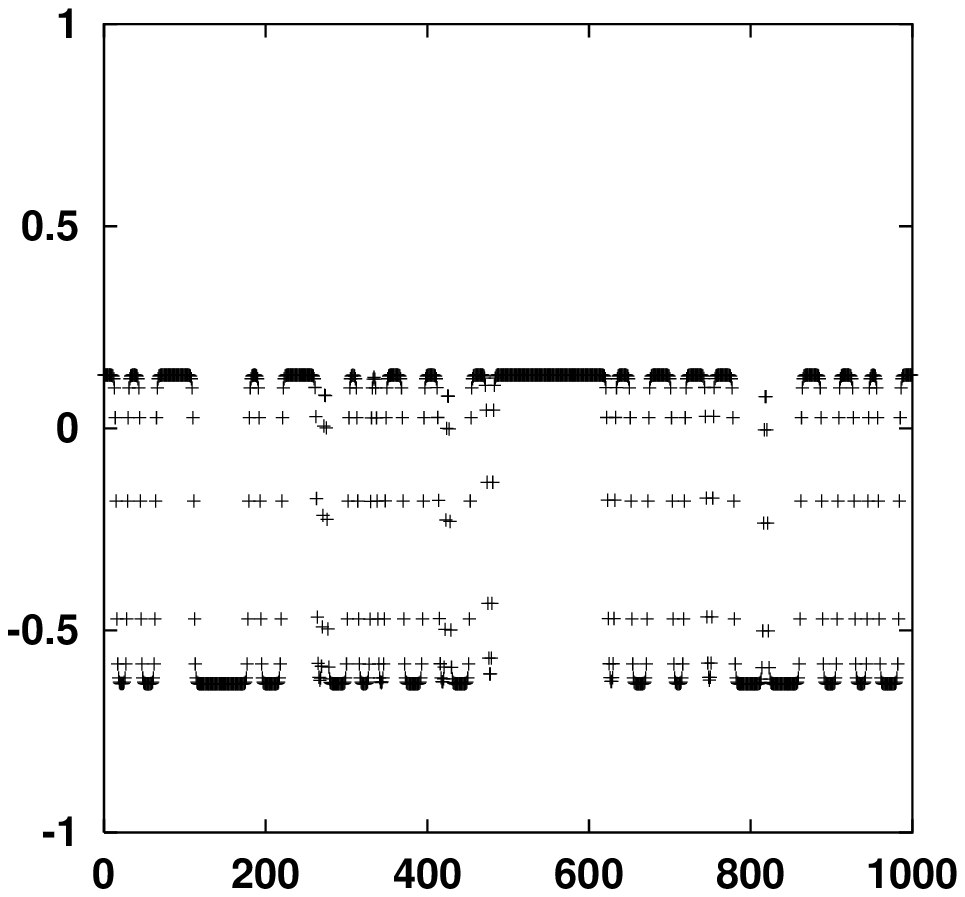}}}
\put(200,220){\scalebox{0.4}{\includegraphics{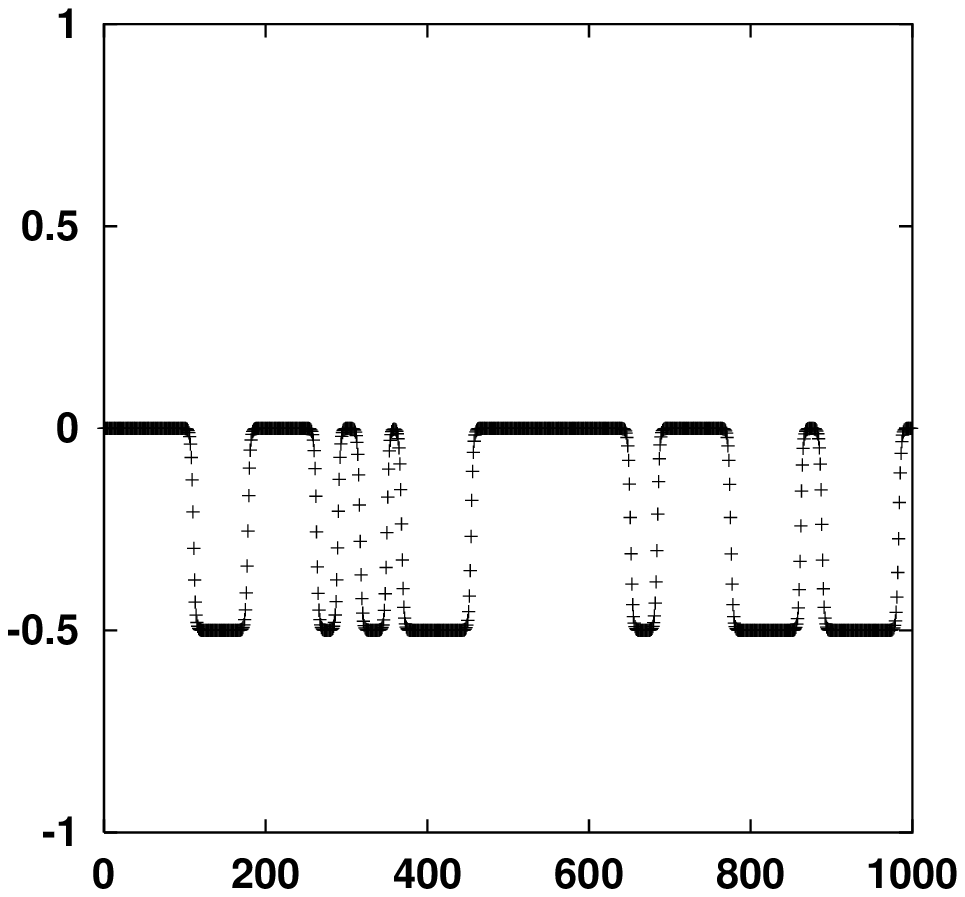}}}
\put(0,20){\scalebox{0.4}{\includegraphics{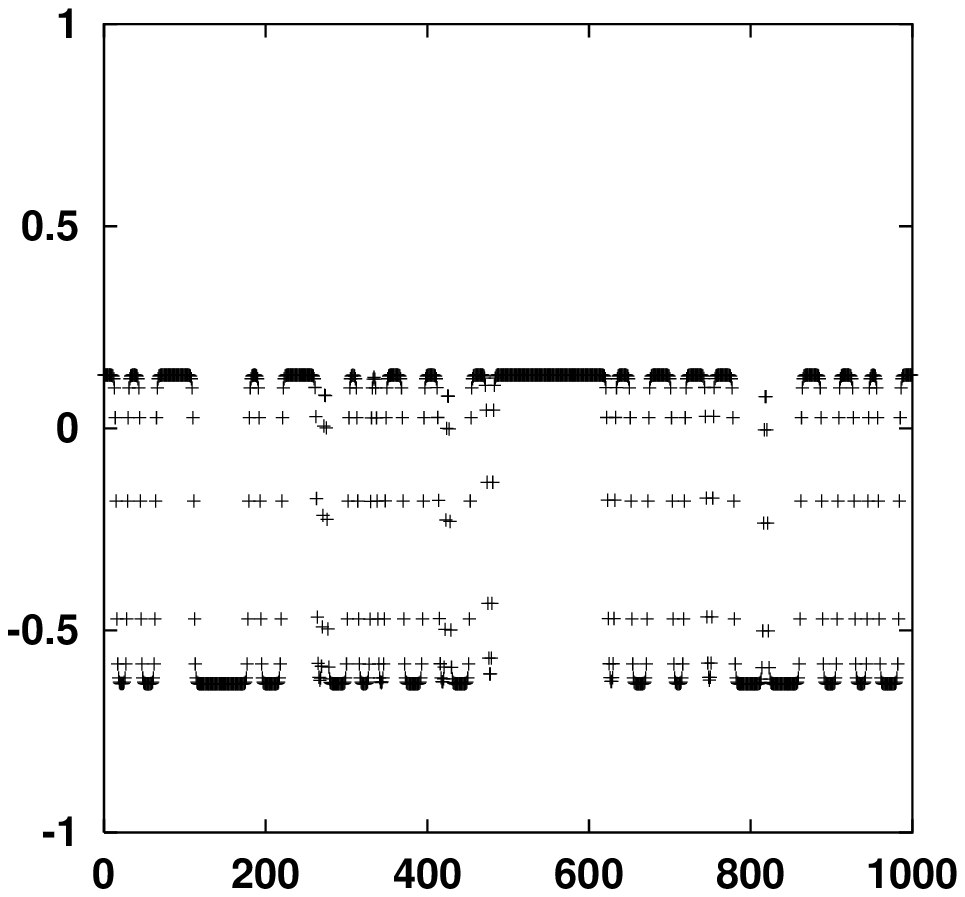}}}
\put(200,20){\scalebox{0.4}{\includegraphics{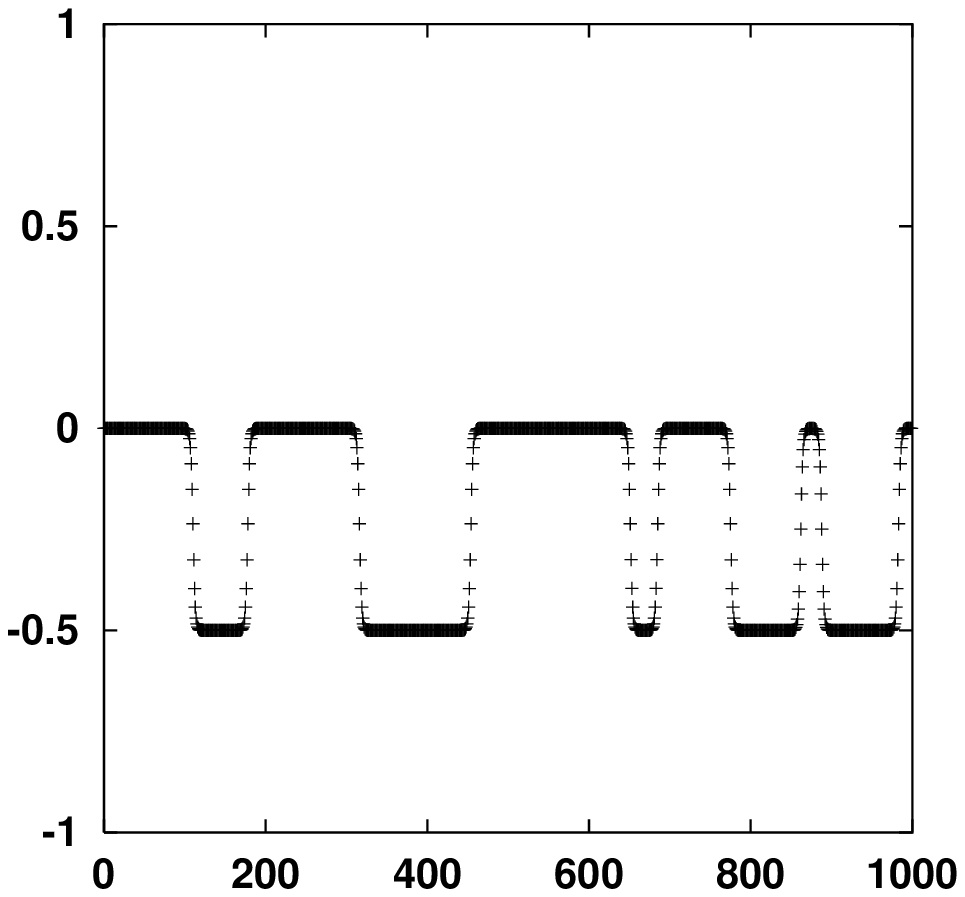}}}
\put(100,0){\Large $x$}
\put(0,100){\Large $\Phi$}
\end{picture}
\caption{Spacelike slices of the period 2 orbit of the $2B_-$ map,
showing frozen disorder ($a=0.4$, left) and exponentially slow relaxation
($a=0.5$, right).  The relaxation times are $10^4$ (top), $10^6$ (middle)
and $10^8$ (bottom)
\label{f:slow}}
\end{figure}

The most interesting question is what has happened to the period 2 orbits.
Looking at the $2B^-$ in figure~\ref{f:non}, the period 2 orbit bounds
a discrete set of values for $0.3675<a<0.48$ and a continuous set of
values for $0.48<a<0.5556$.
In both cases a spatial slice of the solution reveals intermittent switching
between the two values, as shown in Fig.~\ref{f:slow}.
The discrete case contains dynamically ``frozen'' domains which remain for
all time.  The continuous case relaxes by increasing
the size of the domains, but does so exponentially slowly.

In the discrete case, there is no relaxation, so the final state (and
consequently any property of it, such as the average $V$ in Eq,~(\ref{e:V})
depends strongly on the choice of initial conditions; a different distribution
function for the initial conditions would lead to a different distribution of
domains.  Averages can be estimated given a knowledge of the relative
proportions of upper domains, lower domains and boundary regions.  Other
properties such as spatial correlation functions require more information
about the distribution of domains.

The continuous case is numerically ambiguous due to the exponentially long
time scales. It may may relax eventually to the stable periodic orbit, so
its infinite time properties can be calculated directly from this orbit.

As the relaxation time increases, the transition between the discrete
and continuous
regimes (ie $a=0.48$ at a time of $10^4$) is observed to move towards
larger $a$, so it is in fact likely that at sufficiently long times,
the final destination is a state with frozen disorder for all $a$
corresponding to this stable periodic orbit.  For example, the lower right
plot in Fig.~\ref{f:slow} shows that at $a=0.5$ and a time of $10^8$ the
dynamics seems to have frozen.

However, even if as suggested in Ref.~\cite{B01} the coupled map lattice
describes something going on at the Planck scale ($10^{-35}$ seconds),
the transient relaxation may relate to measurable time scales.  Note
that time in the coupled map lattice is not physical time but Parisi-Wu
fictitious time which supposedly is taken to infinity.  The question is
whether this indeed happens or rather it remains exponentially large
but finite.

The other backward coupled ($B$) maps and also $2A_-$ have the same
appearance of multiple ``feathered'' attracting points, which apparently
give way to a continuous distribution as $a$ is varied, specifically
for $2B$ and $a>0.6$, for $2A^-$ and $a\approx 0.3$ or $a\approx 0.7$,
and for $3B$ and $a>0.8$.  It seems likely that this phenomenon of
exponentially slow dynamics and strong dependence on initial conditions
holds for all of these regimes also.

The other general feature, noted in Ref.~\cite{B01}, is that for small $a$,
the advanced ($A$) and backward ($B$) coupled maps lead to very similar
behaviour. 

Comparing Fig.~\ref{f:non} here with the results in Ref.~\cite{B01},
the following points (from which the standard model parameters are calculated)
are in regions with apparently stable attractors:

First, the interaction energy zeros:
$a_3^{(3B)}\approx0.35$ lies in the domain
of stability of a period two orbit, however the numerical results suggest
chaos here (for our uniformly distributed initial conditions), albeit
strongly focussed on a small range of $\Phi$.  $a_1^{(2A)}\approx0.12$ lies
just below the lower boundary of the stable window (of spatial period 2;
not discussed above).  $a_1^{(2A^-)}\approx 0.18$ is close to the boundary
of stability of the period 2 orbit at $1/6$.  $a_2^{(2A^-)}=1/2$ and
$a_2^{(2B^-)}=1$ both lie in the stable (period one) regions as noted
in Ref.~\cite{B01}.

Second, the self energy extrema:
${a'}_7^{3A}\approx 0.17$ and ${a'}_8^{3A}\approx 0.23$ both lie in a
complicated
stable region, as do ${a'}_7^{3B}\approx 0.19$ and ${a'}_8^{(3B)\approx0.29}$,
the latter of which is corresponds to a stable period two orbit.
${a'}_3^{(2A)}\approx0.18$ lies close to the upper boundary of the stable
window mentioned in the previous paragraph.  ${a'}_3^{(2B)}\approx0.22$ lies
in a stable period 2 orbit, although numerical results suggest chaos
in its vicinity.

Thus, these values (and hence standard model parameters according to
Ref.~\cite{B01}) relate to many different types of dynamics, from fully
chaotic (ie not in the lists above) to a stable fixed point.  Only the latter
is easily obtained analytically.

\section{Conclusion}
Coupled Tchebyscheff maps provide excellent scope for both analytic and
numerical studies of stable synchronised states.  Periodic states of
(temporal) length one and two can be found analytically, and are found
to have linear stability for a number of models in various ranges of
the parameter $a$.  Numerical observations confirm transitions at values
predicted by the theory.

There is also a family of orbits of increasing length that can be handled
analytically in the long period limit, which demonstrates the existence
of local stability, even arbitrarily close to the uncoupled fully chaotic
limit $a=0$.

Numerically, the period 2 orbit in the $2B^-$ model is an attractor where
the two values of $\Phi$ occur in different domains at one time slice,
exchanging their values at the next time slice.  The size and distribution
of these domains depends sensitively on the distribution of initial
conditions.  Furthermore, some values of $a$ require exponentially long
times before the final state is reached.  Similar structures appear in the
$2A^-$, $2B$ and $3B$ models. 

How do these results affect Ref.~\cite{B01}?  In
general we might expect the results to depend on the choice of initial
conditions if a stable periodic state exists (true for many of the values)
or indeed any coexistence of attractors (beyond the scope of this paper),
and also be affected by exponentially slow dynamics in one of the ``feather''
regions (apparently not relevant at the values given in Ref.~\cite{B01}).

In addition, the fact that many if not most of the values of $a$
correspond to stability, or at least not full chaos, undermines the
assertion that it is the strong chaotic properties of the Tchebyscheff
maps that is responsible (via stochastic quantisation) for quantum
mechanics.  The coupled Tchebyscheff maps are often far from chaotic,
and some explanation must be given as to why many stable parameter values are
important in particle physics yet the quantisation mechanism requires
strong instability. 

The results presented here are only a brief sketch of the diffusively
coupled Tchebyscheff map lattices, however it has been sufficient to
find analytic results for classes of periodic states and demonstrate some
of the richness of spatially extended dynamics including bifurcations
evident in Figs.~\ref{f:t=2} and~\ref{f:non}, and exponentially
slow dynamics in Fig.~\ref{f:slow}.  The combination of analytic
tractability and a great variety of dynamical behaviour makes the
coupled Tchebyscheff maps good candidates for prototypes
of spatiotemporal chaos, much as the (closely related) logistic map is
a prototype of low dimensional chaos.

\section*{Acknowledgements}
The author thanks C. Beck and J. R. Dorfman for helpful discussions.
Financial support was provided by the Nuffield Foundation, grant
NAL/00353/G.

\end{document}